# The Connection Between Plasmon Decay Dynamics and the Surface Enhanced Raman Spectroscopy Background: Inelastic Scattering from Non-Thermal and Hot Carriers

Shengxiang Wu[†], Oscar Hsu-Cheng Cheng[†], Boqin Zhao[†], Nicki Hogan[†], Annika Lee[†], Dong Hee Son[†,*], Matthew Sheldon[†,‡,*]

[†]Department of Chemistry, Texas A&M University, College Station, TX, 77843, USA

[‡]Department of Material Science and Engineering, Texas A&M University, College Station, TX, 77843, USA

E-mail: sheldonm@tamu.edu

**Abstract**

Recent studies have established that the anti-Stokes Raman signal from plasmonic metal nanostructures can be used to determine the two separate temperatures that characterize carriers inside the metal- the temperature of photoexcited "hot carriers" and carriers that are thermalized with the metal lattice. However, the related signal in the Stokes spectral region has historically impeded surface enhanced Raman spectroscopy (SERS), as the vibrational peaks of adsorbed molecules are always accompanied by the broad background of the metal substrate. The fundamental source of the metal signal, and hence its contribution to the spectrum, has been unclear. Here, we outline a unified theoretical model that describes both the temperature dependent behavior and the broad spectral distribution. We suggest that the majority of the Raman signal is from inelastic scattering directly with carriers in a non-thermal energy distribution that have been excited via damping of the surface plasmon. In addition, a significant spectral component (~ 1%) is due to a sub-population of hot carriers with an energy distribution that is well approximated by an elevated temperature distribution, about 2000K greater than the lattice temperature of the metal. We have performed temperature and power-dependent Raman experiments to show how a simple fitting procedure reveals the plasmon dephasing time, as well as the temperatures of the hot carriers and the metal lattice, in order to correlate these parameters with quantitative Raman analysis of chemical species adsorbed on metal surface.

## I. INTRODUCTION

Recent advances in the study of plasmonic metal nanostructures are enabling new applications in chemical sensing,[1, 2] bio-imaging,[3, 4] light harvesting[5, 6] and non-linear phenomena, such as plasmon-induced magnetism.[7] The primary benefits provided by metal nanostructures are due to the strong subwavelength optical field concentration that results from the coupling of resonant oscillations of free electrons in the metal, termed plasmons, with the incident light field. The strongest field concentration is manifest at 'hot spots' in nanostructures. Therefore, dimers or aggregates of metal nanoparticles are commonly used as plasmonic substrates in sensing applications, and in particular, for surface enhanced Raman spectroscopy (SERS),[8, 9] with plasmonic behavior also enabling related tip-enhanced Raman spectroscopy (TERS) techniques. Although SERS has been studied extensively for over thirty years,[10] the origin of SERS signals continues to be debated, and a variety of mechanisms associated with the SERS effect have been proposed.[11-13]

While it is well established that the signal enhancement (scaling as $|E|^4$) is due to the excitation of plasmonic resonances on the metal substrate,[10] the mechanism underlying light emission or inelastic scattering from plasmonic metals is still unclear.[14-19] On the one hand, it is commonly observed that a bi-exponential decay of the anti-Stokes (aS) signal from plasmonic metals is strongly temperature dependent.[20, 21] Several studies have proposed that the aS spectrum of plasmonic metals can be fitted to extract the lattice temperature of the metal, $T_l$, and the elevated electronic temperature, $T_e$, in the steady state.[22, 23] Recently, the presence of these highly energetic carriers at $T_e$ was probed experimentally,[24-26] and was connected to the plasmon-mediated chemistry.[27, 28] On the other hand, the Raman peaks from molecular vibrational modes observed in SERS spectra are always accompanied by a spectrally broad background on the Stokes side.[13, 16-18] This background signal is present even in the absence of chemical adsorbates (blue trace, Fig. 2d). The physical origin of the broad background has been the subject of much debate, and its presence inhibits quantitative interpretation of SERS molecular spectra.[12, 29, 30] Methods to understand and subtract the background are thus important in analytical SERS applications, and there is little consensus on the best strategy.

Recently, Barnett et al.[16] attributed the SERS background to the interaction of molecular dipoles on metal surfaces with their own image charge. Although this 'molecules in the mirror' approach is able to produce the broad background in the Stokes spectral region, it provides only a partial explanation, since the signal is observed on clean Au without molecular adsorbates.[17] Further, the well-established and pronounced temperature-dependence, especially in the anti-Stokes region of the spectrum, is not accounted for. Alternatively, Lin et al.[13] proposed that the SERS background is due to photoluminescence (PL) from the metal. Similarly, Cai et al.[14, 15] attributed the spectrum from gold nanorods to PL that is enhanced by a Purcell effect. This enhancement is well documented[19] and implicit in the method of Lin et al.[13] as well. However, the SERS background is still observed during excitation with infrared radiation that is not sufficiently energetic to drive vertical interband transitions that preserve momentum during PL, complicating a mechanistic interpretation.[20, 31] Nonetheless, several studies suggest the background signal is, at least partially, the result of an inelastic electronic scattering process intrinsic to the plasmonic substrate itself.[25, 32]

In addition to strong field concentration relevant for sensing applications, resonant plasmonic geometries also produce highly energetic hot carriers that can enable remarkable photocatalytic processes and new strategies for optoelectronic energy generation.[14, 24, 33, 34] Here, we refer to hot carriers as energetic electrons or holes that are not in thermal equilibrium with the lattice phonons of the metal, but that have an energy distribution that can be approximated accurately as a thermal distribution with temperature, $T_e$. The dynamics of hot carriers have been studied extensively using time-resolved ultrafast pump-probe spectroscopy, and experiments have established the following timeline summarized in the well-known two-temperature model (TTM):[35, 36] (1) After optical excitation, a coherent plasmon dephases non-radiatively (10 fs time scale)[37, 38] to generate a non-thermal distribution of excited electron-hole pairs (Fig. 2a). The dephasing time, $\tau_{\text{dephase}}$, is related to materials parameters such as the composition and morphology of the nanostructure, among other factors. Notably, chemical adsorbates and surface chemical reactions couple with the plasmon resonance, modifying the dephasing time in a process termed chemical interface damping (CID).[39-41] (2) Non-thermal carriers then begin to relax through electron-electron scattering (100 fs time scale) to establish an energetic distribution with a characteristic "hot electron" temperature, $T_e$, that is significantly elevated compared with the lattice temperature, $T_l$, of the metal. (3) On the longest time scale, electron-phonon scattering (1-5 ps)[37, 38, 42] results in photothermal heating that increases $T_l$, or is dissipated in the environment. Fundamentally, $T_e$ and $T_l$ depend on the relative magnitude of the electronic and lattice heat capacities of the metal, respectively, so that commonly employed excitation conditions that increase $T_l$ by 10-100's degrees can cause increases in $T_e$ by several thousands of degrees, due to the much smaller electronic heat capacity. Further, the rate of relaxation from the distribution $T_e$ to $T_l$ is proportional to their temperature difference and the electron-phonon coupling constant.

In a series of recent papers,[22-24] we showed there is a clear relationship between the aS Raman spectrum from bare plasmonic substrates and the photoexcited carriers described by the TTM. Specifically, we established how the aS spectrum can be fit quantitatively to determine, $T_e$, $T_l$, and the size of the sub-population of hot carriers, $\alpha$, in the elevated temperature distribution $T_e$ that results during steady-state photoexcitation. Based on the magnitude of $\alpha$, other dynamic information contained in the TTM can also be learned, such as the hot carrier lifetime, $\tau_{e \to ph}$, as well as the rate of coupling to dissipation pathways, e.g. the electron-phonon coupling constant.

In this study we significantly expand the dynamic information can be learned from quantitative analysis and fitting to the steady-state electronic Raman scattering spectrum of plasmonic nanostructures, while additionally providing new insight into the photochemical surface reactions in which they participate. Our primary hypothesis is that the Raman signal across the entire Stokes and anti-Stokes spectral region may be attributed to the relative contribution of inelastic scattering from (1) non-thermal carriers with an energetic distribution dependent on both the plasmon dephasing time, $\tau_{\text{dephase}}$, and the temperature distribution, $T_l$, from which the electrons are photoexcited, and (2) a separate contribution from a subpopulation of hot carriers described by the temperature distribution $T_e$ and population size $\alpha$. Our proposed analysis can subtract the continuum background in SERS studies and correlate the plasmon dephasing behavior with hot carrier induced chemical reactions, due to the clear signature of CID and molecular Raman modes in the spectra. Simultaneously, fitting to our model provides measure of both temperature distributions $T_e$ and $T_l$, and thereby the dynamic information summarized in the TTM, as in our previous reports.

## II. METHODS

### Sample

For our Raman spectroscopy experiments, we fabricated samples of 50 µm×50 µm square arrays of gold nanodisks on a silicon substrate. The detailed fabrication method can be found in ref. 24. In brief, gold nanodisks arrays were prepared with top-down electron-beam lithography (EBL). The gold nanodisk has a diameter of 400 nm and a height of 100 nm at a pitch of 500 nm on a 150 nm thick gold film. Fig. 1 shows a scanning electron microscope (SEM) image and optical image of the fabricated sample, as well as the corresponding reflection spectrum.

### Raman Measurement

Raman spectra and reflection spectra were taken using a confocal microscope (Witec RA300), 50x objective (NA=0.55, WD=8.7mm) and spectrometer (UHTS300, grating = 300 g/mm). For Raman spectra acquisitions, the excitation source is a 532 nm CW Nd:YAG laser with a spot size of 0.55 µm2. The Raman spectrum was normalized to the reflection spectrum, which is proportional to the photonic density of states in our fitting routine. For temperature dependent measurements the Raman spectra were collected on the same microscope system with a heating stage (Linkam TS1500VE).

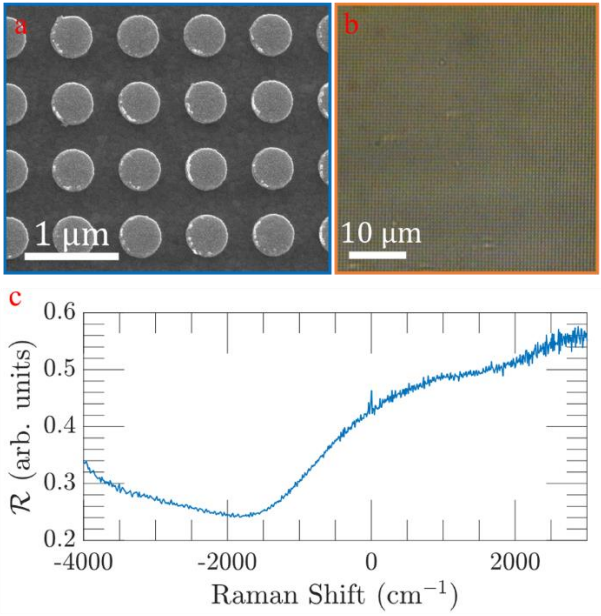

FIG. 1. (a) SEM image and (b) Optical image of EBL-fabricated gold nanostructures, (c) Measured reflection spectrum.

**Non-Thermal Carrier Generation in Plasmonic Nanostructures**

The generation of non-thermal carriers from photoexcitation in plasmonic nanostructures and the corresponding energy distribution, $\Gamma_e$, has been analyzed in several recent theoretical and computational studies.[35, 43-51] Our calculations are based on an adaptation of the theoretical framework (Fermi-Golden rule approach) previously developed by Nordlander et al.[52, 53] and Lischner et al..[54]

$$\Gamma_e(\varepsilon_f, \omega) = \frac{4}{\tau_{\text{dephase}}} \sum_{\varepsilon_f} f(\varepsilon_i)[1 - f(\varepsilon_f)] \left\{ \frac{|M_{fi}(\omega)|^2}{(\hbar\omega - \varepsilon_f + \varepsilon_i)^2 + \hbar^2 \tau_{\text{dephase}}^{-2}} \right. \quad (1)$$
$$\left. + \frac{|M_{if}^*(\omega)|^2}{(\hbar\omega + \varepsilon_f - \varepsilon_i)^2 + \hbar^2 \tau_{\text{dephase}}^{-2}} \right\}$$

Here, $f$ is the Fermi-Dirac distribution function, which for simplicity is commonly assumed to be at zero temperature, (see below for a discussion of how temperature information is incorporated into our calculations), $\tau_{\text{dephase}}$ is the plasmon dephasing time, and $M_{fi} = \int dr V(r,\omega) \rho_{fi}(r)$ is the transition matrix element. Further, $\rho_{fi}(r) = e\Psi_f^*(r)\Psi_i^*(r)$ where $e$ is the elementary charge, subscript i and f stand for initial and final states, and $V(r, \omega)$ is the plasmon-induced potential. It is important here to distinguish the plasmon dephasing time in Eq. 1 with any carrier relaxation time, such as electron-electron or electron-phonon scattering. The dephasing time $\tau_{\text{dephase}}$ defines the natural linewidth of excited plasmon, which is on the femtosecond scale.[55, 56] In Fig. 2a, we plot the calculated non-thermal energy distribution of carriers generated inside a 10 nm thick gold slab that results from dephasing of the surface plasmon. For simplicity, the distribution of non-thermal holes (red bars) is shown with the opposite sign. We note that this energy distribution (Fig. 2a, blue and red bars) can be approximated well using a Lorentzian function (Fig. 2a, solid line), with

$$\mathcal{L} = \frac{\frac{1}{2}\Gamma}{(E - E_\text{F})^2 + \left(\frac{1}{2}\Gamma\right)^2} \qquad (2)$$

where $\Gamma = \frac{\hbar}{\tau_\text{dephase}}$ and $E_\text{F}$ is the Fermi energy of gold. The calculated energy distribution is in good agreement with other computational studies of plasmonic decay phenomena.[46, 48, 51]

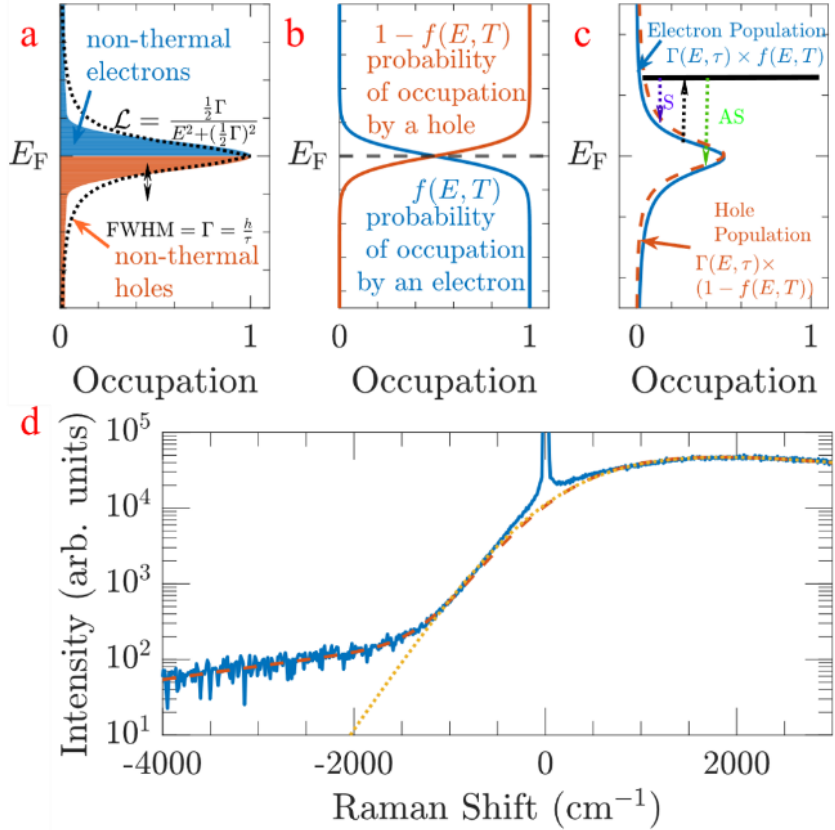

FIG. 2. (a) Calculated non-thermal electron (blue bars) and hole (red bars) populations in a 10 nm gold slab during excitation with 532 nm radiation. This energy distribution can be approximated using a Lorentzian function with energy width determined by the plasmon dephasing time, $\tau_\text{dephase}$(dotted line), as in Eq. 2. (b) Thermal occupation probability for electrons (blue) and holes (red). Multiplication of (a) and (b) yields (c) the joint density of states. (d) Experimentally measured Raman scattering from an array gold nanodisks. The joint density of states (yellow dotted line) deviates from the Raman spectrum at higher aS scattering energy, unless hot carriers at a separate temperature distribution $T_e$ are also accounted for (red dashed line), as in Eq. 4.

**The Joint Density of States**

The electronic Raman signal from nanostructured metals at the Raman-shifted energy, $\hbar\omega$, has been observed to scale with a joint density of states calculation, $J(\hbar\omega)$.[25, 32]

$$J(\hbar\omega) = \int g(E)f(E)g(E + \hbar\omega)\bigl(1 - f(E + \hbar\omega)\bigr)dE \qquad (3)$$

Here, $g(E)$ is an empirically fitted Lorentzian function and $f(E)$ is the Fermi-Dirac distribution at energy $E$, accounting for the thermal activation of carriers in the metal. Eq. 3 provides a description of the intensity across the entire Raman spectrum, since the anti-Stokes signal corresponds to the annihilation of excited carriers separated in energy by $\hbar\omega$ (as opposed to electron-hole pair generation). Fig. 2a-c provides a diagram depicting the electronic Raman signal as being due to inelastic scattering according to $J(\hbar\omega)$ mediated via a virtual transition. We note that this expression does not distinguish if carriers excited at the pump wavelength relax through real or virtual transitions, so that energetic broadening of PL from the metal (i.e. relaxation through real transitions) is also consistent with this model. Further, it should be noted that the physical interpretation of the Lorentzian function $g(E)$ is also unclear. Otto et al.[32] claim that $g(E)$ represents the inhomogeneous electron gas of the metal surface, associated with the surface roughness, while Szczerbinski et al.[25] assume $g(E)$ is the density of states of surface carriers. Thus $g(E)$ is considered to be related to materials properties of the metal.

In contrast, based on our calculation of the non-thermal carriers (Eq. 1) and the comparison with Eq. 2 (Fig. 2a), we hypothesize that $g(E)$, in fact, corresponds to the energy distribution of non-thermal carriers, $\mathcal{L}(E, \tau_{\text{dephase}})$, generated by non-radiative dephasing of the plasmon. According to our proposed microscopic picture, the electronic Raman signal from plasmonic metals is primarily due to inelastic scattering with non-thermal carriers (Fig. 2c) that are generated during the plasmon dephasing. This hypothesis naturally solves a puzzle regarding why the broad background has not been seen from bulk, smooth metal films: nanoscale field localization provides the momentum relaxation required for free space excitation of localized plasmons that then decay into non-thermal carriers.[17]

However, we also emphasize that the proposed scattering mechanism (Eq. 3) is not sufficient to explain Raman signals at more energetic scattering energies in the anti-Stokes region between -4000 cm$^{-1}$ to -1500 cm$^{-1}$. In Fig. 2d we plot the experimental Raman spectrum (blue trace) from arrays of gold nanodisks along with the fit according to Eq. 3 (yellow dotted line), and the deviation is clear. Therefore, as discussed in our previous reports, we also account for the contribution from the sub-population of hot carriers with a characteristic temperature $T_e$. The overall Raman spectrum is then described according to:

$$J(\hbar\omega) = D \cdot \mathcal{R}\left[\int \mathcal{L}(E, \tau_{\text{dephase}}) f(E, T_l) \mathcal{L}(E+\hbar\omega, \tau_{\text{dephase}})(1 - f(E+\hbar\omega, T_l)) dE \right.$$
$$\left. + \alpha \int f(E, T_e)(1 - f(E+\hbar\omega, T_e)) dE\right] \quad (4)$$

where $D$ is a scaling factor that accounts for the collection efficiency of our microscope setup, and $\mathcal{R}$ is the reflection spectrum of the sample which is proportional to the density of photonic states. Analogous scaling factors have been used to fit photoluminescence and anti-Stokes Raman spectra in studies of individual nanostructures.[14, 15] $\mathcal{L}(E, \tau_{\text{dephase}})$ is the energy distribution of non-thermal carriers, which is approximated as a simple Lorentzian function (Eq. 2) in order to simplify the fitting process. The quantity $f(E, T_l)$ is the Fermi-Dirac distribution of carriers thermalized at the lattice temperature, $T_l$, while $f(E, T_e)$ is the Fermi-Dirac distribution of the sub-population of hot carriers at the elevated temperature, $T_e$. The variable $\alpha$ is a free fit parameter that accounts for the different signal intensity, i.e. the relative size, of the population in the hot carrier distribution compared to the population of non-thermal carriers.

To summarize, the majority of the steady-state Raman signal is due to inelastic scattering by non-thermal carriers that have been excited via plasmon dephasing from the bath distribution with temperature $T_l$, while about ~1% of the signal (scaled by $\alpha$) is due to a sub-population of carriers at the elevated thermal distribution $T_e$. The fit to Eq. 4 is excellent and is displayed in Fig. 2d (red dashed line). A more detailed analysis of the fitted trends follows below. For clarity, it is important to note that Eq. 4 analyzes three electronic populations: non-thermal carriers, hot carriers, and carriers thermalized with lattice. These populations are present simultaneously in the steady state. Thus Eq. 4 is comparable to an extended TTM model.[48, 57]

### III.    Results and Discussion

To demonstrate the validity of our physical model, a series of Raman spectra were collected with samples on a temperature-controlled microscope stage during heating from 298 K to 448 K (Fig. 3a). A 532 nm continuous wave (CW) laser at an optical power density of $2.5\times10^9$ W/m$^2$ was used to excite the Raman signal from the gold nanostructures depicted in Fig. 1. In these experiments, we estimate the time interval between two incident photons ( ~1 fs) to be much faster than the time scale of electron relaxation processes (~100s fs). Thus, consecutive photons first excite and then probe the electronic distribution during CW experiments (see SI for details). It is also important to note that our previous experiments indicate a negligible morphology change within the temperature range studied,[24] and in these experiments reproducible Raman spectra, fitted trends, and SEM images before and after the Raman studies rule out the possibility that the major findings illustrated below are due to irreversible structural changes of the nanostructures (see SI for details).

On a logarithmic scale the experimental anti-Stokes trend appears as a bent line, with each region displaying a well-defined slope. We and other authors have previously attributed the region of steeper slope (-700 cm$^{-1}$ to -1200 cm$^{-1}$) to scattering from carriers at the thermal distribution, $T_l$, and the second flatter slope at higher energy (-1200 cm$^{-1}$ to -4000 cm$^{-1}$) to scattering from carriers at the elevated temperature $T_e$.[23-25] Indeed, fitting these separately sloped regions to different Boltzmann distributions indicates temperatures that well approximate the two temperatures obtained through the more robust fitting procedure contained in Eq. 4, and that agree with our independent verification of $T_e$ in thermionic emission experiments.[23, 24]

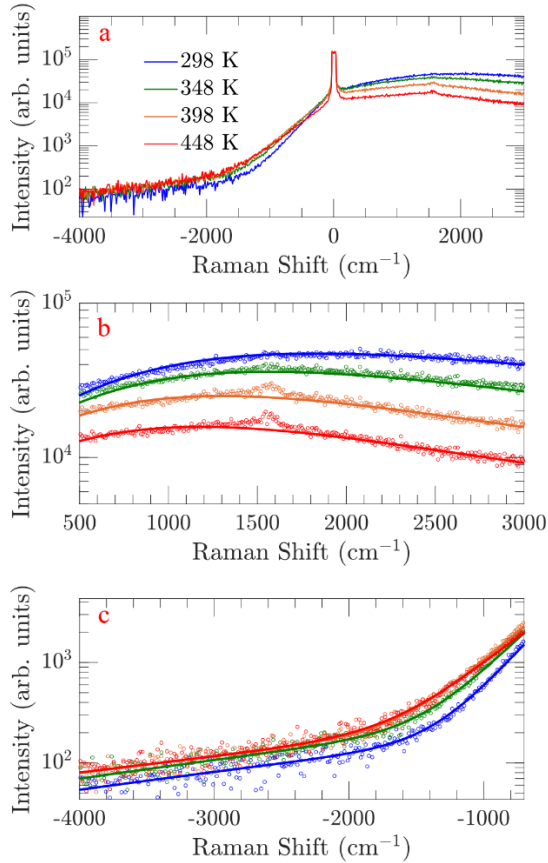

FIG. 3. (a) Raman spectra of gold nanodisks at different stage temperature (dots) and the fit to Eq. 4 (solid lines) (b) Stokes spectrum (c) anti-Stokes spectrum.

Additionally, we observe the appearance of a molecular vibrational signal at 1580 cm$^{-1}$, which is attributed to the G band of a graphite-like material that forms on the sample surface during measurement.[58, 59] This signal is very likely the result of highly energetic hot carriers reducing adsorbed carbon-containing molecules during measurement, and this signal has been observed ubiquitously in SERS.[60] It should be noted that direct coupling between excited plasmons and the molecular orbitals of adsorbates can affect the surface plasmon lifetime (i.e. the dephasing time), depending on the coverage and molecular structure of adsorbates.[40, 41, 61] Therefore, we used Eq. 4 to fit our experimental signal across the entire collected spectral range from -4000 cm$^{-1}$ to 3000 cm$^{-1}$, omitting the Rayleigh line (-200 cm$^{-1}$ to 200 cm$^{-1}$) and the graphitic carbon signal (1200 cm$^{-1}$ to 2000 cm$^{-1}$). The quality of the fit is highlighted for the Stokes side (Fig. 3b) and the anti-Stokes side (Fig. 3c). The fitting expression captures not only the broad signal throughout the Stokes region, but also the two separately sloped regions characteristic of the anti-Stokes signal.

The quantified physical parameters based on these fits are displayed in Fig. 4. The lattice temperature (blue dots, Fig. 4a) is slightly higher than the set stage temperature (blue dashed line) likely due to photothermal heating induced by the laser at the focal spot. The electronic temperature (red dots, Fig. 4b) corroborates well with our previously determined value from thermionic emission measurements[24] as well as other experimental and computational studies.[62] We note that the contribution from hot carriers ($\alpha$) also increases with stage temperature (Fig. 4b). We previously established a relationship between the size of the hot carrier sub-population and the electron-phonon relaxation time with $\tau_{e\rightarrow ph} = \alpha\rho V/N\sigma$, where $\rho$ is the electron density of gold, $V$ is the interacting volume of the laser and the metal surface (we assume an optical penetration depth of ~30 nm in these experiments),[63] $N$ is the number of photons reaching the surface per unit time and $\sigma$ is the experimentally measured absorptivity. That is, in the steady state the size of the population of hot carriers is determined by the ratio of the rate at which carriers are excited and the rate at which they relax. Therefore, the increase of $\alpha$ suggests a slower electron-phonon coupling process with increased temperature (Fig. 4c). A slower rate of hot electron relaxation is consistent with the phonon bath being at an elevated temperature, in accordance with the TTM.

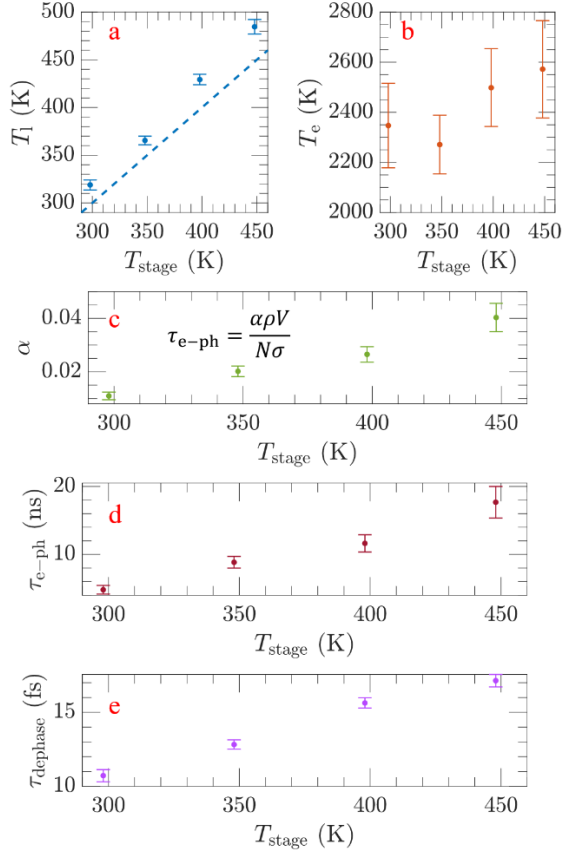

FIG. 4. (a) The lattice temperature (blue dots) is systematically higher than the stage temperature (blue dashed line) due to photothermal heating at the laser spot. (b) The hot carrier temperature (red dots) is comparable to previous studies,[23, 24, 62] (c) The hot carrier contribution, $\alpha$. (d) The electron-phonon relaxation time, $\tau_{e\rightarrow ph}$. (e) The plasmon dephasing time, $\tau_{\text{dephase}}$. The error bars indicate the 95% confidence interval.

Another important feature extracted from fitting is the plasmon dephasing time (Fig. 4d). The fitted dephasing time, $\tau_{\text{dephase}}$, is around 15 fs, and consistent with the reported value for gold.[64] The dephasing time is largely responsible for the spectral shape on the Stokes side of the spectrum, with shorter dephasing time giving rise to a broader energy distribution. We hypothesize that the monotonic increase of the plasmon dephasing time is likely related to the formation of graphitic carbon species on the sample surface during our measurements. Surface chemical species can modify the plasmon dephasing process through CID, which could result in a decrease or increase of the dephasing time based on the nature of the chemical interaction. Based on other reports, our expectation is that CID would cause the plasmon to dephase more quickly,[40, 41] so the opposite trend here could also indicate thermal desorption or chemical transformation of the surface species that caused plasmon damping. The relative contribution of the graphitic carbon signal increased during the course of this experiment.

We performed an additional experiment to provide more insight into the behavior of the dephasing time and its relationship with the observed surface chemical reactions. A separate gold nanodisk array was prepared (200 nm diameter at a pitch of 300 nm) and Raman spectra from the sample were collected using a 532 nm CW laser. The dependence of the Raman signal on the dephasing time is more pronounced in the Stokes-shifted spectrum, so our analysis is focused on the spectral region from -1200 cm$^{-1}$ to 3000 cm$^{-1}$. We first performed a temperature-dependent study by heating the sample stage using the same procedure reported above, but with much lower optical power density ($6.4\times10^7$ W/m$^2$) in order to minimize photothermal heating effects. Next, with the sample stage heater turned off, we analyzed the Raman spectra as a function of incident optical power density. The dependence on stage temperature or optical power is shown in Fig. 5.

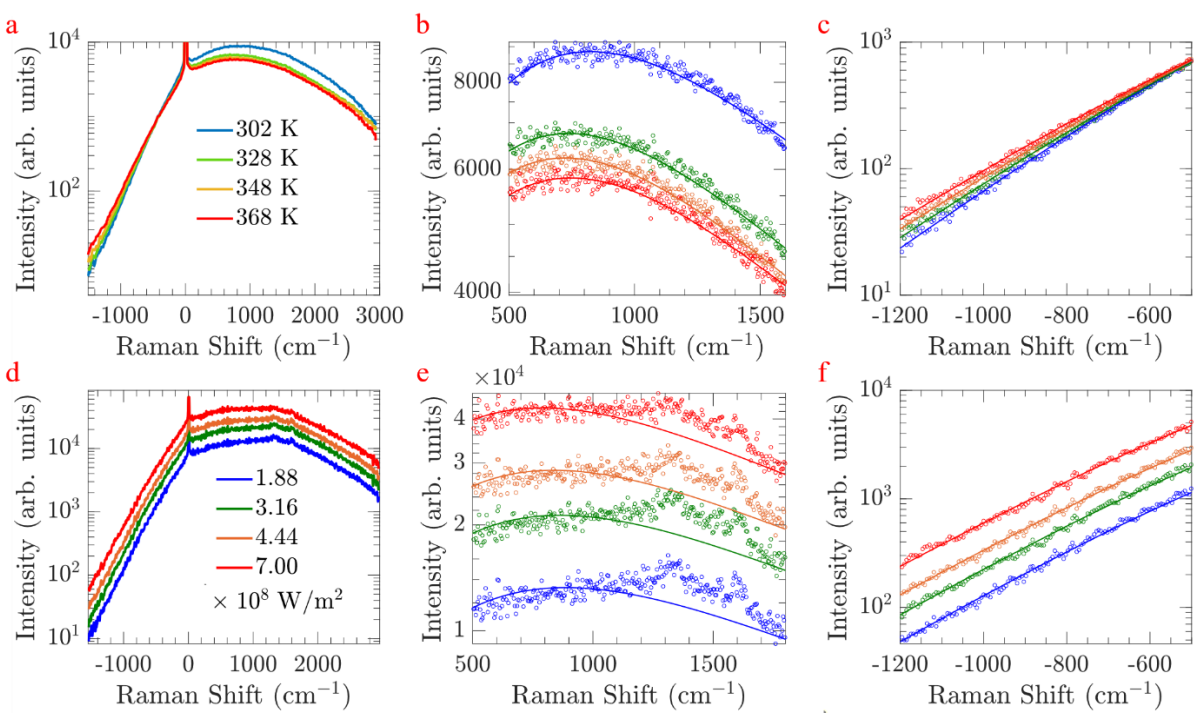

FIG. 5. Raman spectra of gold nanodisk arrays recorded at (a-c) different stage temperature or (d-f) varying incident optical power. (d) Agreement between experiment (dots) and theory (solid lines) across the entire spectral range is highlighted for (b, e) the Stokes region and (c, f) the anti-Stokes region. During fitting, the 1200 cm$^{-1}$ to 2000 cm$^{-1}$ region containing the vibrational peak of the surface carbon species was neglected.

Since the optical power used to collect the temperature-dependent Raman spectra (Fig. 5a) is much weaker than the optical power used in Fig. 5d, we observed no evidence of the formation of graphitic carbon species located at 1350 cm$^{-1}$ (D band) and 1580 cm$^{-1}$ (G band) as shown in Fig. 5b. In contrast, the formation of carbon species during the power dependent study (Fig. 5e) indicates the important role of transiently excited hot carriers for performing surface chemical reactions. Similar behavior is observed by Szczerbinski et al.[25] and in our previous reports. Using the same fitting equation but excluding the contribution from surface carbon, we observed a good agreement between theory and experiment (Fig. 5b-f).

A summary of the fitted parameters is provided in Fig. 6. At low optical power, and thus minimal laser heating, there is very good correlation between the fitted lattice temperature (blue dots) and the heated stage temperature (black dashed line, Fig. 6a), further validating our model. We observed a monotonic increase of lattice temperature when incident optical power was increased (Fig. 6b). For comparisons between experiments, the optical power was modulated so that samples obtained similar lattice temperatures via laser heating or when using the stage heater. The extracted plasmon dephasing time is depicted in Fig. 6c, and is relatively shorter under laser heating. We conclude that the contribution of hot carriers is significant in the optical power dependent study, since the appearance of the D and G band from graphitic carbon is only observed in those spectra (Fig. 5e). The adsorbed carbon species may promote CID and cause the reduced plasmon dephasing time. Although the specific chemistry observed here may be complex, we emphasize that our analysis provides clear spectroscopic signatures that can help distinguish effects due primarily to thermal heating, versus chemical interactions that involve plasmonic phenomena.

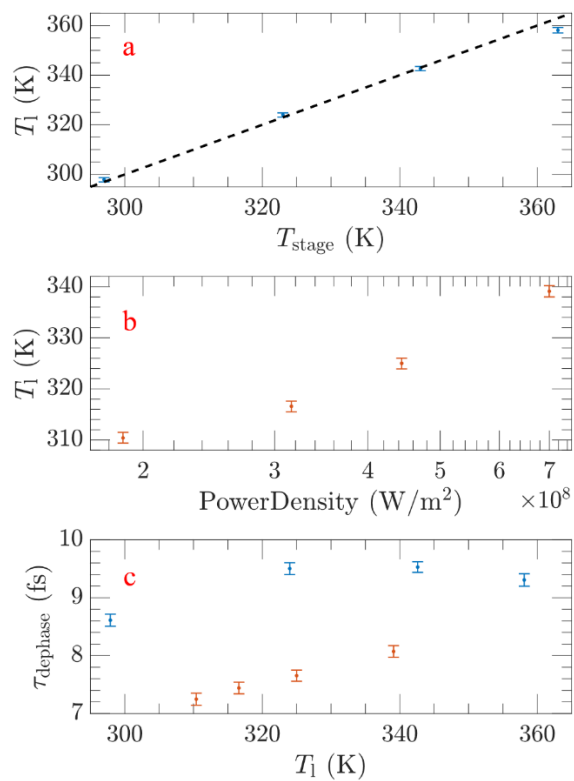

FIG. 6. (a) Correlation between the fitted lattice temperature (blue dots) and the heated stage temperature (black solid line). (b) Extracted lattice temperature (red dots) due to laser heating while the stage was held at room temperature. (c) Comparison of the plasmon dephasing time $\tau_\text{dephase}$ during stage heating (blue dots) or during laser heating (red dots). The error bar represents the 95% confidence interval.

Besides resolving the plasmon dephasing information contained in the steady-state Raman signal, this fitting technique can also quantitatively subtract the continuum background on the Stokes side to reveal the vibrational fingerprint of adsorbed chemical species. This strategy may be useful for monitoring chemical reactions *in situ* during SERS experiments. For instance, by subtracting the fitted curve (solid line, Fig. 5e), the vibrational peaks of the D and G bands of the surface carbon species are easily resolved (Fig. 7a). Based on multi-peak fitting to the molecular vibrations (solid line in Fig. 7a), including analysis of the signal due to sp$^3$ carbon at 1530 cm$^{-1}$, the integrated signal intensity of the carbon vibrational modes with respect to optical power is shown in Fig. 7b (see SI for details). The power-law exponent is less than one, which means the coupling between the carbon vibrational modes and surface plasmon field enhancement decreases with optical power. This decreased coupling, observed by tracking the molecular Raman signal intensity, is corroborated by the observed increase of the dephasing time of the surface plasmon (red dots, Fig. 6f) that is indicated by the spectral shape of the broad Stokes background.

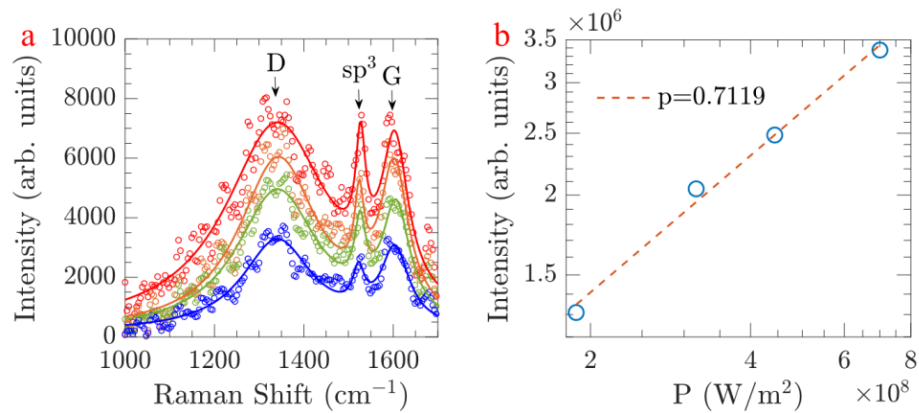

Fig 7. (a) Background-subtracted vibrational peaks of amorphous carbon species on the plasmonic substrate. (b) Integrated signal intensity (solid line in a) versus optical power. The fitted power-law exponent, p=0.7119, is less than one.

## IV. Conclusion

In conclusion, we have developed a theoretical model which hypothesizes that Raman signals from plasmonic nanostructures are due to inelastic scattering with non-thermal carriers generated during plasmon dephasing, and additionally, hot carriers at an elevated temperature greatly in excess of the temperature of the metal lattice. The characteristic temperatures of the TTM and corresponding dynamic information, as well as the surface plasmon dephasing time can be determined quantitatively by fitting a Raman spectrum using our model. We showed excellent correspondence between the fitted lattice temperature of an array of Au nanodisks and the stage temperature on which the sample was heated. Furthermore, a power-dependent laser heating study revealed the insights provided by using our model to correlate the plasmon dephasing information with chemical species adsorbed on the metal surface. We believe these results will inform strategies to monitor chemical reactions using SERS and advance the analysis of plasmon-mediated photocatalytic reactions and other hot carrier effects.

## ACKNOWLEDGEMENT

This work is funded by the Gordon and Betty Moore Foundation through Grant GBMF6882. M.S. also acknowledges support from the Welch Foundation (A-1886) and the National Science Foundation (Grant DMR-2004810).

## DATA AVAILABILITY

The data that support the findings of this study are available from the corresponding author upon reasonable request.

**REFERENCE**

1. Stockman, M. I., Nanoplasmonic sensing and detection. *Science* **2015,** *348* (6232), 287.
2. Larsson, E. M.; Langhammer, C.; Zorić, I.; Kasemo, B., Nanoplasmonic Probes of Catalytic Reactions. *Science* **2009,** *326* (5956), 1091.
3. Vo-Dinh, T.; Wang, H.-N.; Scaffidi, J., Plasmonic nanoprobes for SERS biosensing and bioimaging. *J Biophotonics* **2010,** *3* (1-2), 89-102.
4. Brolo, A. G., Plasmonics for future biosensors. *Nature Photonics* **2012,** *6* (11), 709-713.
5. Green, M. A.; Pillai, S., Harnessing plasmonics for solar cells. *Nature Photonics* **2012,** *6* (3), 130-132.
6. Brongersma, M. L.; Halas, N. J.; Nordlander, P., Plasmon-induced hot carrier science and technology. *Nature Nanotechnology* **2015,** *10* (1), 25-34.
7. Cheng, O. H.-C.; Son, D. H.; Sheldon, M., Light-induced magnetism in plasmonic gold nanoparticles. *Nature Photonics* **2020,** *14* (6), 365-368.
8. Wang, C.-A.; Ho, H.-C.; Hsueh, C.-H., Periodic ZnO-Elevated Gold Dimer Nanostructures for Surface-Enhanced Raman Scattering Applications. *The Journal of Physical Chemistry C* **2018,** *122* (47), 27016-27023.
9. Blaber, M. G.; Schatz, G. C., Extending SERS into the infrared with gold nanosphere dimers. *Chemical Communications* **2011,** *47* (13), 3769-3771.
10. Moskovits, M., Surface-enhanced Raman spectroscopy: a brief retrospective. *Journal of Raman Spectroscopy* **2005,** *36* (6-7), 485-496.
11. Fisher, G. B.; Sexton, B. A., Identification of an Adsorbed Hydroxyl Species on the Pt(111) Surface. *Physical Review Letters* **1980,** *44* (10), 683-686.
12. Burstein, E.; Chen, Y. J.; Chen, C. Y.; Lundquist, S.; Tosatti, E., "Giant" Raman scattering by adsorbed molecules on metal surfaces. *Solid State Communications* **1979,** *29* (8), 567-570.
13. Lin, K.-Q.; Yi, J.; Zhong, J.-H.; Hu, S.; Liu, B.-J.; Liu, J.-Y.; Zong, C.; Lei, Z.-C.; Wang, X.; Aizpurua, J.; Esteban, R.; Ren, B., Plasmonic photoluminescence for recovering native chemical information from surface-enhanced Raman scattering. *Nature Communications* **2017,** *8* (1), 14891.
14. Cai, Y.-Y.; Liu, J. G.; Tauzin, L. J.; Huang, D.; Sung, E.; Zhang, H.; Joplin, A.; Chang, W.-S.; Nordlander, P.; Link, S., Photoluminescence of Gold Nanorods: Purcell Effect Enhanced Emission from Hot Carriers. *ACS Nano* **2018,** *12* (2), 976-985.
15. Cai, Y.-Y.; Sung, E.; Zhang, R.; Tauzin, L. J.; Liu, J. G.; Ostovar, B.; Zhang, Y.; Chang, W.-S.; Nordlander, P.; Link, S., Anti-Stokes Emission from Hot Carriers in Gold Nanorods. *Nano Letters* **2019,** *19* (2), 1067-1073.
16. Barnett, S. M.; Harris, N.; Baumberg, J. J., Molecules in the mirror: how SERS backgrounds arise from the quantum method of images. *Physical Chemistry Chemical Physics* **2014,** *16* (14), 6544-6549.
17. Hugall, J. T.; Baumberg, J. J., Demonstrating Photoluminescence from Au is Electronic Inelastic Light Scattering of a Plasmonic Metal: The Origin of SERS Backgrounds. *Nano Letters* **2015,** *15* (4), 2600-2604.
18. Mahajan, S.; Cole, R. M.; Speed, J. D.; Pelfrey, S. H.; Russell, A. E.; Bartlett, P. N.; Barnett, S. M.; Baumberg, J. J., Understanding the Surface-Enhanced Raman Spectroscopy "Background". *The Journal of Physical Chemistry C* **2010,** *114* (16), 7242-7250.
19. Mertens, J.; Kleemann, M.-E.; Chikkaraddy, R.; Narang, P.; Baumberg, J. J., How Light Is Emitted by Plasmonic Metals. *Nano Letters* **2017,** *17* (4), 2568-2574.
20. Huang, J.; Wang, W.; Murphy, C. J.; Cahill, D. G., Resonant secondary light emission from plasmonic Au nanostructures at high electron temperatures created by pulsed-laser excitation. *Proceedings of the National Academy of Sciences* **2014,** *111* (3), 906.
21. Xie, X.; Cahill, D. G., Thermometry of plasmonic nanostructures by anti-Stokes electronic Raman scattering. *Applied Physics Letters* **2016,** *109* (18), 183104.
22. Hogan, N.; Sheldon, M., Comparing steady state photothermalization dynamics in copper and gold nanostructures. *The Journal of Chemical Physics* **2020,** *152* (6), 061101.
13